\documentstyle[twocolumn,prl,aps]{revtex}
\tighten
\begin{document}
\title{Why do Thin Polymer Films Melt at Lower Temperature? A Continuum
Approach}
\author{R. Seemann, K. Jacobs, and S. Herminghaus}
\address{Dept. of Applied Physics, University of Ulm, D-89069 Ulm,
Germany}
\date{\today}
\maketitle

\begin{abstract}

We have investigated the reduction of the glass transition
temperature, $T_g$, for thin supported films of particularly small
molecular weight ($M_W$ = 2 kg/mol) polystyrene, and found good
agreement with earlier studies on larger molecules. By combining
the eigenmodes of the films with mode coupling theory, we arrive
at a simple and predictive model which is in quantitative
agreement with the data. Its {\it only fitting parameter} is the
shear modulus governing the relevant fluctuations. The weak
dependence of the latter upon the molecular weight of the polymer
is in accordance with the general observation that the reduction
of $T_g$ is largely independent of chain length at sufficiently
low molecular weight.
\end{abstract}

The experimental observation that thin polymer films melt at
temperatures well below the glass transition temperature of the
bulk polymer \cite{KJ1} is one of the major current challenges in
the theory of homopolymers. Although one might at first glance
expect such a behaviour, due to the impact of the finite size
geometry of a thin film upon objects as large as polymer
molecules, a slightly closer look reveals the intricacy of the
problem: the effect of a noticeable reduction of the glass
temperature $T_g$, namely, occurs in many cases at film
thicknesses orders of magnitude larger than the end-to-end
distance, $R_{ee}$, of the molecules. A fundamental understanding
of this effect would be of great interest not only for polymer
physics, but potentially also elucidate the physics of the glass
transition in general.

For the sake of clarity, but without loss of sufficient
generality, let us restrict our discussion to the case of
polystyrene (PS), which is among the best studied polymers in this
context. As it has recently been pointed out in very thorough and
elucidating studies \cite{MFB,DVF}, it is necessary to distinguish
between long chain polymers with molecular weight above about 380
kg/mol, and shorter ones. For long PS chains, with a molecular
weight of, say, 500 kg/mol, a substantial reduction is observed at
a thickness below approximately 60 nm. Although $R_{ee}$ is still
less than that, the length of a stretched molecule of this weight
is of the order of one micron. In this case, the effect might
indeed be attributed to finite size effects upon the chain
conformation and chain motion. These ideas have been meanwhile
developed and seem to reproduce successfully the features
experimentally observed \cite{DeGennes,DVDG}.

For shorter chains, however, the puzzle is more involved, since it
is meanwhile accepted that the characteristic dependence $T_g(h)$
upon the thickness is largely independent of the chain length
\cite{DVF,Forrest} in this case. A finite size effect upon the
geometry and motion of the single chain is thus ruled out as the
basic mechanism. Many attempts have been undertaken to explain
these findings on other grounds, mostly by considering microscopic
models of the inner structure of the films. D. Long and F. Lequeux
have envisaged the freezing of the film as a percolation of rigid
domains \cite{Moukarzel}, mediated by thermal fluctuations
\cite{LongLequeux}. Other models assume a layered structure of the
film, with a particularly mobile region close to the free surface
of the film. Within this framework, Forrest and Mattsson
\cite{Forrest} have recently been able to achieve quite impressive
accordance with the experimental data \cite{MFB,Forrest}. Their
model makes use of the so-called cooperativity length, $\xi(T)$,
which plays a mayor role in a whole class of theoretical concepts
of the glass transition. The only drawback is that there is yet no
well established theory of $\xi(T)$. Furthermore, as a consequence
of the two-layer structure of the film inherent in the model, it
is not completely clear why there should not be two glass
transitions rather than a single one shifted in temperature.

As a complementary approach, computer simulations of polymer films
with free surfaces have recently been carried out, and most of the
experimental findings were obtained qualitatively \cite{Torres}.
However, the polymer chains in these simulations were very much
shorter than those used in the experiments (16 monomer units),
well below the entanglement limit. It is therefore worthwile to
study experimentally the behaviour of polymers with short chain
length, both to ease comparison with simulation and to explore the
range of validity of the apparent independence of the shift of
$T_g$ on molecular weight.

We have thus investigated the glass transition in thin supported
films of PS with a molecular weight as small as 2 kg/mol ($\approx
20$ monomer units). The polymer was purchased from Polymer Labs
(UK) with a polydispersity index $M_W/M_N$ = 1.05, the radius
of gyration is 1.3 nm. Effects from the molecular geometry are
thus expected only for films of few nanometers thickness. The
films were spin cast from toluene solution onto silicon wafers
(Silchem GmbH, Freiberg/Germany), which were previously cleaned by
ultrasonication in acetone, ethanol, and toluene, subsequently.
Residual organics were removed with Caro's acid (1:1 $H_2SO_4 +
H_2O_2$), and the substrates were thoroughly rinsed with hot
millipore water afterwards. Films were investigated with
thicknesses ranging from 4 to 160 nm. The roughness of the free
surface of the films was less than 0.2 nm, as revealed by scanning
force microscopy (SFM).

The glass transition temperature was determined in several ways
independently. The standard procedure of monitoring the thermal
expansion of the film via ellipsometry, as introduced by Keddy and
Jones \cite{KJ1}, was used for film thicknesses down to 9.6 nm.
Thermal expansion coefficients agreed well with previously
published data both above and below $T_g$. Possible loss of
material during the heating and cooling procedures was checked by
applying the Clausius-Mosotti relation to the thickness and
refractive index data obtained from the ellipsometry. Perfect
conservation of material was invariably found. At large film
thickness, $T_g$ obtained in this way approached $327 \pm 1$ K,
which is consistent with the temperature at which macroscopic
melting is observed in the bulk ($T_g^{0}$) for PS with this chain
length.

For thinner films, we observed the buildup of amplified thermal
fluctuations (spinodal dewetting) \cite{Karim,OurPRL}. These
processes were monitored by SFM with {\it in-situ} heating. In
order to speed up the experiments to a feasible time scale,
dewetting was observed at temperatures close to $T_g^{0}$. In this
way, the viscosity $\eta(T)$ was determined. Furthermore, we found
that $\eta(T)$ varied with temperature according to a
Vogel-Fulcher law, just as the bulk viscosity, but with the
Vogel-Fulcher temperature shifted by a certain amount depending on
film thickness. In this way, the apparent glass transition
temperature of the film  could be obtained by extrapolation.

Our data are shown in fig.~\ref{Tg(h)} as the full symbols. The
circles represent the thermal expansion measurements, the squares
were obtained by extrapolation from spinodal dewetting experients.
As one can clearly see, the glass transition temperature is
substantially reduced for all films thinner than about 50 nm. The
solid line represents the function
\begin{equation}
T_g = T_g^0(1+h_0/h)^{-1}
\label{KimFormel}
\end{equation}
This form has been shown before to account well for the data
obtained by others for larger molecular weight films, if $h_0 =
0.68$ nm was assumed for PS \cite{Kim}. Within experimental
scattering, our data exhibit indeed the same dependence of
$T_g(h)$ in the full range of film thickness explored. Although
our polymer chains are roughly by a factor of 50 shorter, we
obtain $h_0 = 0.82$ from the fit, which is quite close to the
above value. This confirms the weak dependence (if there is a
significant one at all) of the reduction of the glass transition
temperature on the molecular weight of the polymer, down to a
molecular weight as small as 2 kg/mol, comparable to what has been
studied by computer simulation \cite{Torres}.

It is clear that the behaviour displayed in fig.~\ref{Tg(h)} can
in no way be attributed to the geometrical impact of the finite
film thickness upon the microscopic conformation of the individual
chains. Were therefore present here a novel approach to the
problem which intentionally makes no reference to the molecular
structure of the film. Inspired by the `melting' of the spinodal
waves, we consider instead the eigenmode spectrum of the
(viscoelastic) polymer film. This can be considered with all
possible boundary conditions at the substrate, such that films
with large extrapolation length, or free standing films, may as
well be treated within the same framework.

The spectrum can be obtained in a straightforward manner by
combining standard theory of elasticity \cite{LL1} and
hydrodynamics \cite{LL2} in the limit of small Reynolds number
(Stokes dynamics). The equation of motion reads
\begin{equation}
\lbrace \partial_t + \omega_0 + \frac{E}{\eta} \rbrace \nabla^2
{\bf \phi} = \frac{\nabla p}{\eta}
\label{EoM}
\end{equation}
where $E$ is Young's modulus, $\eta$ is the viscosity, and
$\omega_0$ is the rate of relaxation of the individual chains by
diffusion. ${\bf \phi}$ is a vector field related to the strain
tensor, ${\bf S}$. If we restrict ourselves to one lateral ($x$)
and one normal ($z$) coordinate, ${\bf \phi}$ is defined via
\begin{equation}
{\bf S} = \left(\begin{array}{cc} \partial_x \phi_x & \frac{1}{2}
(\partial_x \phi_z + \partial_z \phi_x) \\
\frac{1}{2}(\partial_x\phi_z + \partial_z\phi_x) &
\partial_z\phi_z
\end{array}\right) \label{strain}
\end{equation}
Finally, $p$ is the pressure field.

For harmonic excursions of the free surface, $\zeta(x) =
\zeta_0 \exp\{iqx-\omega t\}$, eq.(\ref{EoM}) has solutions of the type
\begin{eqnarray}
\phi_x = [1+(h+q^{-1})\alpha(q)]\cosh qz - q^{-1}\alpha(q)\sinh qz
\nonumber \\ \phi_z = [1+h\alpha(q)]\sinh qz - z\alpha(q)\cosh qz
\label{solution}
\end{eqnarray}
where for the function $\alpha(q)$, we find
\begin{equation}
\alpha(q) = \left(\frac{q}{2}\right) \frac{e^{2qh}-1}{e^{2qh}-1+qh}
\label{alphafree}
\end{equation}
for free standing films (symmetric modes) as well as for supported
films with full slippage. At the free surface, we used the
boundary conditions of zero tangential stress and $p =
-\sigma\partial_{xx}\zeta$, where $\sigma$ is the surface tension
of the polymer. For free standing films, $h$ is defined here as
{\it half} the film thickness, such that the form of
eqs.~(\ref{solution}) and (\ref{alphafree}) accords with the
experimental observation that the shift of $T_g$ for a free film
of thickness $2h$ is roughly the same as for a supported film of
thickness $h$ \cite{footnote}.

For the relaxation rates of the modes, we get
\begin{eqnarray}
2\omega = (\omega_0 + \frac{E}{\eta} + \frac{\sigma q^2}{2 \eta
\alpha(q)}) \nonumber \\
\pm \sqrt{(\omega_0 + \frac{E}{\eta} + \frac{\sigma q^2}{2 \eta
\alpha(q)})^2 + \omega_0^2\frac{2 \sigma q^2}{\eta \alpha(q)}}
\label{dispersion}
\end{eqnarray}
As it will become clear below, it is mostly the high frequency
modes which contribute appreciably to the reduction of the glass
transition temperature. We thus consider only the high frequency
branch of eq.~(\ref{dispersion}). Since $\omega_0 <<
\frac{E}{\eta}$, the latter is approximately given by
\begin{equation}
\omega = \omega_0 + \frac{E}{\eta} + \frac{\sigma q^2}{2 \eta
\alpha(q)}
\end{equation}
which is monotonously increasing with $q$. However, modes with $q$
much larger than the inverse film thickness, $h^{-1}$, do not
penetrate appreciably into the film, such that only a small
fraction of the material takes part in these modes. Hence we are
led to considering chiefly those modes for which $q \approx
h^{-1}$, since these are the highest frequency modes comprising
more or less all of the film material.

The central assumption is now that the physical cause for the melting, or freezing,
of these thin film modes are {\it memory effects} in the polymer material. These are
of course not included in the above linear theory, and may be formulated in a generic
way by means of a suitable memory kernel, which may be written as
$m\{\phi(t)\} = a_1 \phi + a_2 \phi^2 + a_3 \phi^3 + ...$ \cite{Goetze1,Goetze2}.
It should be well noted that we are considering the strain field, $\phi$, which 
describes the local state of the material. The equation of motion (2) is of first 
order in time, hence we have
\begin{equation}
\phi^{\prime} + \omega\phi + \lambda\int_0^t
m\{\phi(\tau)\}\phi^{\prime}(t-\tau)d\tau\ = 0 \label{MCT}
\label{MCTrelax}
\end{equation}
where $\lambda$ is a coupling parameter. This type of mode coupling equations has been
thoroughly analysed \cite{Leutheusser} in relation
to the {\it microscopic} physics of the glass transition (to which we do not refer here),
as well as to large scale degrees of freedom \cite{Lequeux}. It was found
that the coefficients $a_i$ of the memory kernel vary concurrently with temperature,
and that upon crossing a certain border in the space spanned by the $a_i$, the system
freezes into a nonergodic state \cite{Goetze2}. The existence of such a freezing
transition has been
found to be largely independent of the precise form of the memory kernel.
For the simple scaling analysis to be done here, we therefore simply assume the $a_i$
to vary all in the same way, which appears to be justified closely enough to the
transition. In what follows, we thus assume that all temperature dependence may be
absorbed into the coupling parameter, $\lambda$, and the system freezes when a  
critical value, $\lambda_c$, is reached.

From eq.~\ref{MCTrelax} we see directly that $\lambda$ has the
same dimension as $\omega$, and there is no further independent
time scale in the system. Since we expect $\lambda \rightarrow 0$
for $T \rightarrow \infty$, and $\lambda \rightarrow \infty$ for
$T \rightarrow 0$, we can expand $\lambda = b_1 \frac{T_c}{T} +
b_2 (\frac{T_c}{T})^2 + ...$, with $\Sigma b_i = \lambda_c$.
Retaining only the leading term \cite{footnote2}, and setting $q =
1/h$ as discussed above, we arrive at a simple formula for the
glass transition temperature:
\begin{equation}
T_g(D) = T^0_g\left(1+\frac{1.16 \
\sigma}{h(E+\eta\omega_0)}\right)^{-1} \label{result}
\end{equation}
This is exactly the form of eq.~(\ref{KimFormel}), and fig.~1
shows that this describes our data very well. There was no
significant improvement by including higher terms in the expansion
series for $\lambda$.

The quantity
$\eta\omega_0$ is on the order of a few kPa and can in general be
safely neglected against $E$. The characteristic length scale
which appears here, and which was called $h_0$ in eq.~(\ref{KimFormel}),
is thus the ratio $\sigma/E$, with the surface tension $\sigma =
31$ mN/m for PS. It should thus be well noted that the elastic modulus
determining the dynamics of the relevant modes, $E$, is the only physical fitting
parameter in the model.

The physical picture which emerges from this model is that as the
temperature is increased, melting proceeds as the fastest mode
involving all of the film material detaches from its freezed state
and fluctuates. The concomitantly increased motion of polymer
chains reduces the effective viscosity also for modes with smaller
$q$, which, as a consequence, are sped up and melt in turn. In
this way, the film finally melts at all length scales. One might
apply this view also to the surface of a bulk polymer sample, for
which the spectrum is obtained setting $\alpha = q/2$. As it is
readily seen, one should expect surface melting at the polymer
surface, down to a thickness of $h_{sm} = h_0
T_g^{0}(T_g^{0}-T)^{-1}$. To the best of our knowledge, there is
yet no conclusive evidence in favour or in disfavour of polymer
surface melting. This may be left to further studies.

Let us turn back to our experimental data. From $h_0 = 0.82$ nm,
as obtained from the fit, we find for the elastic modulus $E =
44.1$ MPa, which is, on logarithmic scale, right in between the
modulus of the frozen material (a few GPa) and the modulus just
above $T_g$ (about 300 kPa). As it is well known, both values
display no marked dependence on the molecular weight, such that on
the basis of our model, $T_g(h)$ is expected as well to be largely
independent of molecular weight. This is in agreement with all
data collected so far for molecular weights up to about 300
kg/mol. Nevertheless, the exact physical significance of this
somewhat arbitrary choice of $E$, which may be viewed as $E$ {\it
at} $T_g$, is to be investigated in further studies.

\acknowledgements The authors owe many very helpful hints to J. A.
Forrest and K. Dalnoki-Veress. We are furthermore indebted to D.
Johannsmann, J. Baschnagel and R. Blossey for stimulating
discussions. Funding from the Deutsche Forschungsgemeinschaft
within the Priority Program `Wetting and Structure Formation at
Interfaces' is gratefully acknowledged.

\begin{figure}
\caption{The glass transition of thin films of 2 kg/mol
polystyrene, as determined from thermal expansion (circles) and
from the growth of spinodal waves (squares). The solid curve represents
our model, which has the elastic modulus governing the dominant
modes as its only fitting parameter. It furthermore corresponds to
what was found for larger molecular weight PS before
\protect\cite{Kim}. Top: linear scale. Bottom: logarithmic scale,
showing more details at small film thickness.} \label{Tg(h)}
\end{figure}


\begin{references}
\bibitem{footnote} For different boundary
conditions, the precise form of the solution changes, and it will be interesting to
study whether this corresponds to the observed dependence of the shift of $T_g$ upon
the chemical composition of the substrate. However, since the extrapolation length
was not characterized in these experiments, this aspect will not be pursued here.
\bibitem{footnote2} This form is particularly well justified for memory effects in polymers,
as it will be shown in a forthcoming paper.
\end{references}
\end{document}